\documentstyle[12pt, fleqn]{article}
\def\be{\begin{equation}}
\def\ee{\end{equation}}
\def\ba{\begin{eqnarray}}
\def\ea{\end{eqnarray}}

\def\fun#1#2{\lower3. 6pt\vbox{\baselineskip0pt\lineskip.
9pt\ialign{$\mathsurround=0pt#1\hfill##\hfil$\crcr#2\crcr\sim\crcr}}}

\def\m{{\cal M}}

\def\ApJ{{\it Astrophys. ~J. } }
\def\numu{{\nu}_{\mu}}
\def\nuee{{\nu}_e}
\begin{document}
\begin{titlepage}
\null\vspace{-62pt}
\begin{flushright} DOE--ER--40682--87\\
CMU--HEP--94--33 \\
To appear Phys. Rev. D52, 1995\\
\end{flushright}
\vspace{0.5in}
\centerline{\large \bf{A small non-vanishing cosmological constant}}
\centerline{\large \bf{from vacuum energy:physically and observationally
desirable}}
\vspace{.5in}
\centerline{Anupam Singh}
\vspace{.4in}
\centerline{{\it Physics Department,  Carnegie Mellon University,
Pittsburgh PA~~ 15213}}
\vspace{.5in}
\baselineskip=24pt
\begin{quotation}

Increasing improvements in the independent determinations of the
Hubble constant and the age of the universe now seem to indicate that
we need a small non-vanishing cosmological constant to make the two
independent observations consistent with each other.
The cosmological constant can be physically interpreted as due to the
vacuum energy of quantized fields.  To make the cosmological observations
consistent with each other we would need a vacuum energy density,
$ \rho_v \sim (10^{-3} eV)^4 $ today
( in the cosmological units $ \hbar=c=k=1 $ ).
It is argued in this article that such
a vacuum energy density is natural in the context of phase transitions
linked to massive neutrinos.
In fact, the neutrino masses required to
provide the right vacuum energy scale to
remove the age Vs Hubble constant discrepancy
are consistent with those required to solve the solar neutrino problem
by the MSW mechanism.
\end{quotation}
\end{titlepage}
\newpage

\baselineskip=24pt

\vspace{24pt}

\section{Introduction}

Increasing accuracy in astronomical observations is leading us to an
increasing precision in the determination of cosmological parameters.
This in turn is leading us to critically re-examine our cosmological models.
In particular,  the precise determination of the Hubble constant and
the independent determination of the age of the universe is forcing us to
critically re-examine the simplest and most appealing cosmological model
- a flat universe with a zero cosmological constant\cite{Pierce,Freedman}.

The Hubble constant enters in the relationship between the recession velocity
of an object and it's distance from us.  The recession velocity of an object
can be determined by using the Doppler effect and is relatively easy to
determine.  It is the calibration of the extragalactic distance ladder
which is the difficult part of measuring the Hubble constant and in which the
precision has been increasing significantly.  It is the use of the
Cepheid variables as standard candles that has allowed the improved
determination of the extragalactic distance scale.  Cepheids are variable
stars whose pulsation period are very stongly correlated with their
luminosities.  These stars are well understood theoretically and the period -
luminosity relationship is well-documented empirically.
By observationally determining the pulsation period of a Cepheid variable and
using the period luminosity relationship one can immediately determine the
luminosity of the object.  Then by using the apparent brightness of the
object one can accurately determine the distance to the Cepheid.
Pierce et al\cite{Pierce} have used this technique
with ground-based observations to determine the
extragalactic distance scale.  An excellent discussion on this subject is
contained in the article by Pierce et al\cite{Pierce} and references therein.

 Pierce et al\cite{Pierce} have determined the Hubble constant to be,
$ H_o = 87 \pm 7 km~s^{-1}~Mpc^{-1} $.  They further point out in their
article that this value of the Hubble constant is in fact in conflict with the
independent determination of the age of the universe\cite{DAVandenBerg}
using Galactic globular
clusters if we use the standard cosmological model with a zero
cosmological constant.  The estimate of the age derived from an analysis of the
galactic globular clusters is $ 16. 5 \pm 2 ~ Gyr $.

An accurate determination of the Hubble constant is also one of the
important goals of the Hubble Space Telescope (HST).
In fact, Freedman et al \cite{Freedman} have used the HST to
calibrate the extragalactic distance scale and hence determine the
Hubble constant.
They obtainn the value of the Hubble constant to be
$80 \pm 17 km~s^{-1}~Mpc^{-1}$. They also point out that their
determination of the Hubble constant is inconsistent with the
age of the globular clusters within the framework of standard
$\Omega = 1$ cosmology with no cosmological constant.

\section{Resolution of the age Vs.  Hubble constant problem through the
introduction of a small vacuum energy. }

One of the ways to avoid the apparent conflict between the observed age
of the universe and the observed Hubble constant is to introduce a small
cosmological constant in the Einstein equations that govern the evolution
of the universe.
This idea has been extensively studied by a number of people including
Tayler\cite{Tayler} and Klapdor and Grotz\cite{KG}.

Let's quickly summarize how a cosmological constant of the right magnitude
can solve the apprent conflict between the age and the Hubble constant.
This can be seen from the
following analysis\cite{rockymike}.
To a good approximation our universe is spatially homogenous and isotropic on
large scales.  It is therefore appropriate to describe space-time by
Robertson-Walker metric which can be written in the form,  ( units $ c = 1 $ )

\be
ds^2 = dt^2 - R^2(t) \left[ \frac{dr^2}{1 - k r^2} + r^2 d\theta^{2} + r^2
\sin^{2}\theta d\phi^{2} \right]
\ee

where,  $ (t, r, \theta, \phi) $
are the comoving coordinates describing a space time point and $R(t)$ is the
 cosmic scale factor.   Also,  $ k= +1,  -1$ or $ 0$ depending on when the
universe
is closed,  open or flat.

The present expansion age of a matter dominated universe can be
evaluated in a Robertson-Walker universe.
General fine-tuning arguments as well as the
inflationary picture gives us a preference for a flat universe,
$ \Omega_o = 1$.  In this case,  $ t_o = \frac{2}{3} H_o^{-1} $.  For,
$ \Omega_o \simeq 1 $,  one can expand the above expressions in a Taylor
expansion,

\be
t_o = \frac{2}{3} H_o^{-1} \left[ 1 - \frac{1}{5} (\Omega_o - 1) + . . .
\right]
\ee

We can also determine the present age of the universe containing
both matter and vacuum energy such that $ \Omega_{vac} + \Omega_{matter} = 1$,

\be
t_o = \frac{2}{3} H_o^{-1} \Omega_{vac}^{-1/2} \ln \left[ \frac{1 +
\Omega_{vac}^{1/2}}{(1 - \Omega_{vac})^{1/2}} \right]
\ee

This will give us much longer lifetimes as can be seen most dramatically
by examining the limit $ \Omega_{vac} \rightarrow 1 $ in which case
$ t_o \rightarrow \infty $.
Indeed having an  $ \Omega_{vac} \sim 0.8 $ is one of the possible
solutions of the age Vs. Hubble constant discrepancy as can be seen
from the following discussion.

First, let us quickly recall the observational numbers on the Hubble constant
and the ages of globular clusters. Here it is worthwile pointing out that
there are actually 2 sets of numbers which though consistent with each other
have slightly different central values and error estimates.
Pierce et al\cite{Pierce} quote the result of their analysis as yielding
a Hubble constant, $H_o$ of $87 \pm 7 km~s^{-1}~Mpc^{-1}$.
They then draw attention
to the fact that this is in conflict with the age estimate of the
globular clusters inferred by VandenBergh's\cite{DAVandenBerg} analysis which
gives an age of $16.5 \pm 2 ~Gyr$.

Freedman et al\cite{Freedman} have a greater amount of data that they have
analyzed very thoroughly. Thus they found over 20 Cepheids in a Virgo
cluster galaxy as opposed to 3 Cepheids found by Pierce et al.
They obtain a slightly different central value of $H_o$ with substantially
larger and perhaps more realistic error bars than Pierce et al.
They thus quote a Hubble constant value,
$H_o = 80 \pm 17  km~s^{-1}~Mpc^{-1}$.
Also based on a wider spectrum of data for ages they quote a central
value of the age of the universe as $14~ ~Gyr$ with $1 ~ \sigma$ error
bars of $ \pm 2~ ~Gyr$.
Freedman et al, however, too point out that even with their
more generous error bars there is still a discrepancy between the
Hubble constant and the age of the universe if we restrict ourselves
to a standard $\Omega = 1 $ cosmology with a zero cosmological constant.

Let us now see how a non-zero cosmological constant can solve this discrepancy
and what values of the cosmological constant are typically required
to minimally solve the discrepancy between the Hubble constant and
the age of the universe. First, let us consider the lowest possible value
of the Hubble constant as quoted by Freedman et al which is
$63~ km~s^{-1}~Mpc^{-1}$. Let us consider this together with the
age of the universe, $t_o = 14.5 ~Gyr$ which though slightly higher than
the central value quoted by Freedman et al is at the lower limit of the
age quoted by Pierce et al. These two values of $H_o$ and $t_o$
imply a value of $H_o~t_o = 0.93$.
This corresponds to the value of $\Omega_{vac} = 0.66$.
Thus, $\Omega_{vac} = 0.66$ would remove the contradiction between the Hubble
constant as determined by Freedman et al and the age of globular
clusters.
However, let us consider a few more values of $\Omega_{vac},~H_o$ and
$t_o$ to see where future more precise observations and analysis might lead
us.
Consider $\Omega_{vac} = 0.7$ which implies $H_o~t_o = 0.964$.
For $H_o = 63 km~s^{-1}~Mpc^{-1}$ this would imply the age $t_o = 15 ~Gyr$
and for $t_o = 14.5 ~Gyr$ this would imply a $H_o = 65  km~s^{-1}~Mpc^{-1}$.
Similarly, $\Omega_{vac} = 0.8$ would imply $H_o~t_o = 1.076$ which
would give an $H_o = 73 km~s^{-1}~Mpc^{-1}$ for $t_o = 14.5 ~Gyr$ and
for a $t_o = 16.5 ~Gyr$ would imply $H_o = 64  km~s^{-1}~Mpc^{-1}$.
Finally, if we consider the lower bound implied by the
numbers quoted by Pierce et al, viz. $H_o = 80 km~s^{-1}~Mpc^{-1}$ and
$t_o = 14.5 ~Gyr$, we get  $\Omega_{vac} \sim 0.85$.
Clearly however one would not like to push the values of $\Omega_{vac} $
much higher than this number for a  number of reasons.
First, we would like $\Omega_{vac} + \Omega_{matter} = 1$ and higher
values of $\Omega_{vac}$ will start to conflict with the lower bound
on matter density from galaxies and clusters. Furthermore, one can start
to place an independent constraint on the cosmological constant
fromm gravitational lens statistics.
Thus based on the HST Snapshot survey of quasars Maoz and Rix\cite{MaozRix}
have inferred the bound $\Omega_{vac} \leq 0.7$.
Fukugita, Hogan and Peebles\cite{FHP} have stated that this constraint
may be relaxed slightly if one considers the fact that at large distances
typical of lensing events galaxies are observed to have larger star
formation rates than their nearby counterparts
and may also contain larger quantities of dust. This may lead to some
lensing events getting obscured and hence the upper limit for
$\Omega_{vac}$ may rise if this is taken into account.
It is unclear how much the upper limit will be pushed upwards
but the fact remains that for both the above reasons one would
not like an extremely large value of $\Omega_{vac}$ even though
a value of $\Omega_{vac} \sim 0.8$ perhaps best meets all the
observational constraints outlined above.
Improved more accurate results and more exhaustive analysis
on all these fronts will clearly shed further light on all these
connected issues.
Let us now turn our attention to a discussion of the possible
physical origin of such an $\Omega_{vac}$.

The value $ \Omega_{vac} \sim 0. 8 $ corresponds to an energy density of the
vacuum
energy density of $\rho_v \sim  (10^{-47} GeV^4) $. This energy density is of
course
much lower than most familiar energy scales in particle physics and the
question
naturally arises as to the physical origin of this energy scale. The smallness
of this
energy scale has been frequently referred to as the cosmological constant
problem.

However,  we'll argue in this
paper that in fact a cosmological constant of the right magnitude required to
 make the cosmological observations consistent with each other may follow
from the dynamical evolution of our universe.  The basic physical picture
which will allow us to arrive to this conclusion is that the cosmological
constant might be interpreted as the vacuum energy of the quantized fields.

This point has been made by many people and is discussed at length by Birell
and Davies\cite{BD}.
Further,  this vacuum energy is not a static quantity but a function of time.
This idea too has been extensively explored by a number of people including
Peebles and Ratra\cite{PeebRa}, Freese, Adams, Frieman and Mottola\cite{FAFM},
Reuter and Wetterich\cite{rwet}.  In fact,  we know that there were
a number of  phase transitions in the evolution
of the universe.

Thus,  the history of the universe  may be summarized as periods of dramatic
change charecterized usually by phase transitions with relatively quiet
 periods of relaxation between the phase transitions \cite{rockymike}.
Indeed,
since the vacuum energy density changes as the
$ ( characteristic~~ energy~~ scale )^4 $
at a phase transition,  in the absence of fine-tuning one expects that the
vacuum energy density at the end of a phase transition
$ \sim ( characteristic~~ energy~~ scale )^4$

This idea has been spelt in detail in the paper by Wilczek\cite{Wilczek}
and also by Reuter
and Weterich\cite{rwet}.  Thus,  $ \rho_v \sim (10^{15}~  GeV)^4$ at the Grand
Unified Symmetry breaking,  $ \rho_v \sim (10^{2}~  GeV)^4$ at the elctroweak
symmetry breaking and  $ \rho_v \sim (10^{-1}~  GeV)^4$ at the
chiral symmetry breaking in QCD.

Furthermore,  at the conclusion of a phase transition the vacuum energy starts
 decaying more slowly to the energy scale characterized by next phase
transition.  This point of view is implicit in the papers by
Wilczek\cite{Wilczek}
and has
been explicitly stated by Reuter and Wetterich.
In fact,  the physical mechanism
 for the decay of vacuum energy is coupling to lighter fields.  This mechanism
 is briefly discussed by Freese, Adams, Frieman and Mottola\cite{FAFM},
who do an extensive analysis of
cosmology with a decaying vacuum energy.

Thus,  the question of the magnitude of the cosmological constant really
becomes a question about energy scales.  Almost every paper on the subject
of the cosmological constant has had to struggle with the characteristic
energy scale of $ \sim 10^{-3} eV$ in the form
$\rho_v \sim ( 0. 003 eV )^4$
in various guises such as $ (10^{-47} GeV^4 ) $ or
$ \Lambda / M_{pl}^2 \sim 10^{-120} $

The fact that $ 0. 003 eV $ is so much less than any characteristic energy
which familiar to most of us has caused a great deal of consternation.

	However,  the energy scale 0. 003eV is certainly not a complete stranger
 to us.  The most natural low-energy scale that  particle physics gives us
 is the light neutrino masses that follow from the see-saw model of neutrino
 masses.  In fact,  the neutrino masses required to to solve the solar
neutrino problem by MSW mechanism\cite{MSWtheory}
imply neutrino masses $ \sim 10^{-3} eV $

This,  of course,  is a powerful hint but is not yet a solution to the
cosmological   constant problem.

In fact,  the finite temperature behaviour of the see-saw model of neutrino
masses has  been studied in detail by Holman and Singh\cite{HolSing}.
The original motivation
for studying this model  was to provide a concrete particle physics model
fo the Late Time Phae Transitions model for structure formation.  Our analysis
showed that in fact this model does exhibit a phase transition with a critical
temperature $ T_c \sim  (few)~~  m_\nu $.

\section{Late Time Phase Transitions and the time evolution of the Hubble
parameter}

In this section we will discuss the cosmological motivations and particle
physics
models for Late Time Phase Transitions.  Once we have a specific model we will
study the
time evolution of fields and the scale factor in this model.  In particular,
we will
be interested in studying the time evolution of the Hubble parameter and will
see that
in this model the Hubble `` constant ``,  in fact has an acceptable value at
the present
age of the universe.

Phase transitions that occur after the decoupling of matter and radiation have
been discussed in the literature as Late Time Phase Transitions (LTPT's). The
original motivation for considering LTPTs\cite{HillFry} \cite{WASS} \cite{PRS}
\cite{FrHW} \cite{GHHK}  \cite{HolSing}was the need to reconcile the
extreme isotropy of the Cosmic Microwave Background Radiation (CMBR)\cite{CMBR}
with the existence of large scale structure\cite{LSS} and also the existence of
quasars at high redshifts\cite{quas}.

Discussions of realistic particle physics models capable of generating LTPT's
have been carried out by several authors\cite{GHHK}  \cite{FrHW}. It has been
pointed
out that the most natural class of models in which to realise the idea of
LTPT's are models of neutrino masses with Pseudo Nambu Goldstone Bosons
(PNGB's). The reason for this is that the mass scales associated with such
models can be related to the neutrino masses, while any tuning that needs to be
done is protected from radiative corrections by the symmetry that gave rise
to the Nambu-Goldstone modes\cite{'thooft}.

Holman and Singh\cite{HolSing} studied the finite temperature behaviour of the
see-saw model of neutrino masses and found phase transitions in this model
which
result in the formation of topological defects.  In fact, the critical
temperature in this model is naturally linked to the neutrino masses.

The original
motivation for studying the finite temperature behaviour of the see-saw
model of neutrino masses came from a desire to find realistic particle physics
models for Late Time Phase Transitions. It now appears that this may also
provide a physically appealing and observationally desirable magnitude for
the cosmological constant.

In particle physics one of the standard ways of generating neutrino masses has
been the see-saw mechanism \cite{ramslan}. These models involve leptons
and Higgs fields interacting by a Yukawa type interaction. We computed the
finite temperature effective potential of the Higgs fields in this model. An
examination of the manifold of degenerate vacua at different temperatures
allowed us to describe the phase transition and the nature of the topological
defects formed.

To investigate in detail the finite temperature behaviour of the see-saw
model we selected a very specific and extremely simplified version of the
general see-saw model. However,
we expect some of the qualitative features displayed by our specific
simplified model to be at least as rich as those present in more
complicated models.

\subsection{ A particle physics model for LTPT}

We chose to study the $2$-family neutrino model. Because of the mass hierarchy
and small neutrino mixings \cite{MSWtheory} we hope to capture some of the
essential physics of the $\nuee$-$\numu$ system in this way. The $2$-family
see-saw model we consider requires 2 right handed neutrinos $N_R^i$ which
transform as the fundamental of a global $SU_R(2) $ symmetry. This symmetry is
implemented in the right handed Majorana mass term by the introduction of a
Higgs field $\sigma_{ij}$, transforming as a symmetric rank $2$ tensor under
$SU_R(2)$ (both $N_R^i$ and $\sigma_{ij}$ are singlets under the standard
model gauge group). The spontaneous breaking of $SU_R(2)$ via the vacuum
expectation value (VEV) of $\sigma$ gives rise to the large right handed
Majorana masses required for the see-saw mechanism to work. Also, the
spontaneous
breaking of $SU_R(2)$ to $U(1)$ gives rise to 2  Nambu Goldstone Bosons.
The $SU_R(2)$ symmetry is explicitly broken in the Dirac sector of
the neutrino mass matrix, since the standard lepton doublets $l_L$ and the
Higgs doublet $\Phi$ are singlets under $SU_R(2)$. It is this explicit
breaking that gives rise to the potential for the Nambu Goldstone modes
via radiative corrections due to fermion loops. Thus, these modes become Pseudo
Nambu Goldstone Bosons (PNGB's).

The relevant Yukawa
couplings in the leptonic sector are:
\be
-{\cal L}_{\rm{yuk}} = y_{ai} \bar{l_L}^a N_R^i \Phi + y \overline{N_R^i}
N_R^{j\ c}
\sigma_{ij} + \rm{h. c. }
\ee
where $a,  i,  j = 1,  2$.  The $SU_R(2)$ symmetry is implemented as follows:
\ba
N_R^i & \rightarrow & U^i_j\ N_R^j \nonumber \\
\sigma_{ij} & \rightarrow & U^k_i \sigma_{kl} (U^T)^l_j
\ea
where $U^i_j$ is an $SU_R(2)$ matrix.
The first (Dirac) term above transforms as an
$SU_R(2)$ doublet,  thus breaking the symmetry explicitly.

We now choose the VEV of $\sigma$ to take the form\cite{LI}:
$\langle \sigma_{ij} \rangle = f \delta_{ij}$,  thus breaking $SU_R(2)$
spontaneously down to the $U(1)$ generated by $\tau_2$ (where $\tau_i$ are the
Pauli matrices).  We take $f$ to be much larger than the doublet expectation
value $v$.

We can parametrize $\sigma_{ij}$ so as to exhibit the Nambu-Goldstone modes as
follows:
\ba
\sigma(x) & = & U(x) \langle \sigma \rangle U^T (x) \nonumber \\
 & = & f U(x) U^T(x)
\ea
with $U(x) = \exp(i(\xi_1 \tau_1 + \xi_3 \tau_3)\slash f)$ (note that $U$ is
symmetric).

After the Higgs doublet acquires its VEV,  we have the following mass terms for
the neutrino fields:
\be
-{\cal L}_{\rm{mass}} = m_{ai} \bar{\nu}_L^a N_R^i + M \overline{N}_R UU^T
N_R^c + \rm{h. c. }
\ee
where $\nu_L^a$ are the standard neutrinos,  $m_{ai} = y_{ai}\ v/{\sqrt{2}}$,
$M = y f/\sqrt{2}$.

Diagonalizing the neutrino mass matrix in the standard see-saw approximation
($|m_{ai}|<<M$) and performing a chiral rotation to eliminate the $\gamma_5$
terms,  we find that the $\xi_i$ dependent {\em light} neutrino masses are
given
by
\ba
m_1^2 & = & \frac{1}{M^2} [\cos^2 2||\xi||\ (m_{11}^2 + m_{12}^2)^2 +
\sin^2 2||\xi||\ (\widehat{\xi_3}
(m_{11}^2 - m_{12}^2) + 2 m_{11} m_{12} \widehat{\xi_1})^2]\nonumber \\
m_2^2 & = & \frac{1}{M^2} [\cos^2 2 ||\xi||\ (m_{21}^2 + m_{22}^2)^2 +
\sin^2 2||\xi||\ (\widehat{\xi_3}
(m_{21}^2 - m_{22}^2) + 2 m_{21} m_{22} \widehat{\xi_1})^2]\nonumber \\
\ea
where $||\xi|| = \sqrt{\xi_1^2 + \xi_3^2}/f$ and
$\widehat{\xi_i} = \xi_i\slash \sqrt{\xi_1^2 + \xi_3^2}$.
We neglect the effects of the heavier neutrinos since they will be suppressed
by powers of $m_{ai}/M$ in the loops that will generate the effective potential
for the $\xi_i$'s.

For simplicity,  to begin with we shall restrict ourselves
to the case where the Dirac mass matrix $m_{ai}$ is
proportional to the identity: $m_{ai} = m \ \delta_{ai}$.  We will consider a
more general case  later.
Using standard results
on the computation of the effective potential due to fermion loops\cite{CW}
(we treat the
$\xi_i$'s as classical background fields,  i. e.  we do not allow them to
propagate in loops),  we can calculate the one-loop
effective potential for the $\xi_i$'s.  The renormalized
potential can be expressed as follows:
\be
V_{\rm{renorm}}(\xi_1,  \xi_3) = V_0 + m_r^2 {\cal{M}}^2 + \lambda_r
({\cal{M}}^2)^2
-\frac{1}{8 \pi^2} (\m^2)^2 (\ln \frac{{\cal{M}}^2}{\mu^2} - \frac{1}{2})
\ee
where $\mu$ is the subtraction point.  Note that we can absorb the effects of
$\lambda_r$ by redefining $\mu$.
We will suppose that this has been done in what follows.
Further, the quantity
${\cal{M}}^2$, is given by

\be
{\m }^2 = {\frac{m^4}{M^2}} (\cos^2 2 ||\xi|| +
\widehat{\xi}_3^2 \sin^2 2 ||\xi||).
\ee

The finite temperature correction due to the two
light neutrinos is given by,
\be
\Delta V_T(\xi_1, \xi_3) = -4 \frac{T^4}{\pi^2} \int_0^{\infty} dx x^2 \log
\left[1 + \exp - \left[ x^2 + \frac{{\cal{M}}^2}{T^2}\right]^{1/2} \right]
\ee
The above expression can be evaluated numerically for any given choice
of parameters,  however for some purposes it is useful to expand the
above expression to get analytic expressions.
Performing the high temperature expansion of the complete potential
and discarding terms of order $(\m^2)^3\slash T^2$ or higher we get,
\be
V_{\rm tot} (\xi_1, \xi_3) = V(\m^2) = (V_0 - \frac{7 \pi^2 T^4}{90}) +
(m_r^2 + T^2/6)\m^2 + \frac{(\m^2)^2}{8 \pi^2}
(n - \log \frac{T^2}{\mu^2} ),
\ee
where $n = 2 \gamma - 1 -2 \log \pi \sim -2. 1303$, $m_r$ is a parameter in the
model and $\mu$ is the renormalisation scale. $\m$ is naturally of the neutrino
mass scale in this model.

A study of the manifold of degenerate vacua of the effective potential at
different temperatures revealed phase transitions in this model accompanied
by the formation of topological defects at a temperature of a few times the
relevant neutrino mass . Typically at higher temperatures the manifold of
degenerate vacua consisted of a set of disconnected points whereas at lower
temperatures the manifold was a set of connected circles.
Thus, domain walls would
form at higher temperatures which would evolve into cosmic strings at lower
temperatures.

Since the critical temperature of the phase transition,
$ T_c \sim (few)~~ m_{\nu} $ ,  let us quickly summarise the observational
evidence for a small $ m_{\nu} $.

At neutrino detectors around the world,  fewer electron neutrinos are
received from the sun than predicted by the Standard Solar Model.
An explanation of the deficiency is offered by the MSW
mechanism\cite{MSWtheory}
which allows the $\nuee$ produced in solar nuclear reactions to change into
$\numu$. This phenomenon of neutrino mixing requires massive neutrinos with the
masses for the different generations different from each other
\cite{MSWtheory}.

The model we considered earlier was an extremely simple one. Although it had 2
families of light neutrinos,  there was only one single light neutrino mass. As
such this model was not compatible with the MSW effect. However it is fairly
straightforward to modify our original model to make it compatible with the MSW
effect as is shown in what follows.

To ensure that it is
not possible to choose the  weak interaction eigenstates to coincide with
the mass eigenstates we must require the 2
neutrino mass scales to be different. We can ensure neutrino mixing in our
model
by demanding that $m_{ai}$ be such that $m_{11} \ne m_{22}$ and
$m_{12}=0=m_{21}$.
In this case, the effective potential
$V_{\rm tot} (\xi_1,  \xi_3) = 1/2 (  V( \m _1 ^2) + V( \m _2 ^2) )$ with
$ V( \m _i ^2)$ having the same
functional form as $ V( \m^2) (i=1, 2)$ and $\m _i^2$ given by the following
expression:
\be
{\m _i}^2 = {\frac{m_{ii}^4}{M^2}} (\cos^2 2 ||\xi|| +
\widehat{\xi}_3^2 \sin^2 2 ||\xi||)
\ee.

Further,  if $m_{11}<<m_{22}$  then $V_{\rm tot}(\xi_1,  \xi_3) = V(\m
_2^2)/2$,
which is exactly half the finite temperature effective potential we discussed
earlier except the neutrino mass scale is the heavier neutrino mass scale.
Hence,
the discussion on phase transitions and formation of topological defects we
carried out earlier goes through exactly except that the critical temperature
is
determined by the mass scale of the heavier of the 2 neutrinos.

In the complete picture of neutrino masses\cite{MSWtheory}, the neutrinos might
have a mass hierarchy analogous to those of other fermions. Further, we expect
that the mixing between the first and third generation might be particularly
small .
In this
scheme,  it is a good first approximation to consider $2$-family mixing. We are
here particularly interested in the $\nuee$-$\numu$ mixing. This is also the
mixing to which the solar neutrino experiments are most sensitive. A complete
exploration of MSW solutions to the solar neutrino problem has recently been
reported by Shi, Schramm and Bahcall\cite{MSWexpt}.
We shall restrict ourselves to the $2$-family mixing.
The data seems to imply a central value for the mass of the muon
neutrino to be a few meV\cite{Bludman}.

We now turn to a quick discussion of the distortions of the
Cosmic Microwave Background Radiation (CMBR)
this model produces.
The most significant microwave distortion
comes from collapsing domain wall bubbles.  This has been discussed
and calculated by Turner,  Watkins and Widrow (TWW)\cite{TurnWW}.
 As pointed out by TWW this anisotropy is most significant
on $\sim 1^o $ angular scales.  The temperature shift due to a photon
traversing a collapsing domain wall bubble is
\be
\frac{\Delta T}{T} = 2. 64 \times 10^{-4} h^{-1} \beta A \sigma / (10 MeV^3)
\ee
where h, A, $\beta$ are dimensionless numerical constants of order unity
and $\sigma$ is the surface tension of the domain wall.  The present
measurements of the CMBR anisotropy then imply\cite{sclu} that
$\sigma < 0. 5 MeV^3 $.

An estimate of
$\sigma$ in terms of the quantities $m_{\nu}$ and $f$ introduced in our model
can be obtained\cite{Widr}. (To make contact with the work of L. Widrow
cited above please note that his $\lambda m^4 = m(\numu)$ and $m = f$
in our notation. ) Thus,  the constraint on $\sigma$ then implies that
$f < 10^{15} GeV$.  Our model is clearly an effective theory with $f$
being some higher symmetry breaking scale on which it is tough to get
an experimental handle.  However,  the constraint derived above is in fact
natural in the context of the see-saw model of neutrino masses embedded
in Grand Unified Theories as discussed by Mohapatra and Parida (MP)\cite{Moha}
and also by Deshpande, Keith and Pal(DKP)\cite{Desh}.

\subsection{Time evolution in the LTPT model}

Now that we know the potential in which the fields $\xi_1$ and $\xi_3$ evolve,
we can write down the coupled set of evolution equations  which describe the
time evolution of the fields and the scale factor of the universe.  Once again
we'll follow the general techniques described in Kolb and
Turner\cite{rockymike}.
The time evolution of the scale factor is given by equations like
(2), (3) and (4).
It is perhaps worth noting that the expression for the presssure and
energy density of the fields is given by:
\be
\rho_{\xi} = \frac{\dot \xi_1 ^2 + \dot \xi_3 ^2}{2} + V(\xi_1, \xi_3)
\ee
\be
p_{\xi} = \frac{\dot \xi_1 ^2 + \dot \xi_3 ^2}{2} -  V(\xi_1, \xi_3)
\ee

The time evolution of the fields  $\xi_1$ and $\xi_3$
in an FRW universe is given by,
\be
\ddot \xi_1 + 3 \frac{\dot R(t)}{R(t)} \dot \xi_1 + \frac{\partial V}{\partial
\xi_1} = 0
\ee
\be
\ddot \xi_3 + 3 \frac{\dot R(t)}{R(t)} \dot \xi_3 + \frac{\partial V}{\partial
\xi_3} = 0
\ee

These coupled equations describing
the time evolution can be solved numerically.
Here we are interested in the time evolution at very recent
epochs. Clearly, for extremely recent epochs the high
temperature expansion is inappropriate.
However, the zero temperature potential which we have computed
and described is a good approximation for stuudying
the time evolution at recent epochs.
Thus we will use the zero temperature potential to do the time
evolution in what follows.
In fact as it turns out the time evolution of
the scale factor is fairly insensitive
to the initial conditions on the fields
but is determined primarily by the order of
magnitude of the energy density in the fields.
In fact we evolved the system with a
variety of initial conditions on the fields and
observed an almost identical time
evolution for the scale factor.

We have investigated the behaviour of the system for a variety
of choice of parameters entering into the potential
with the typical order of magnitude for the parameters
$m_r \sim m_{\nu}$ and $V_o \sim m_{\nu}^4$.
Since it is the fact that the vacuum energy density
in our model is $ \sim  m_{\nu}^4$ that plays a crucial
role in the time evolution in our model, let us re-emphasise
why this is natural in the context of our model.

Recall,  that the original
$-{\cal L}_{\rm{yuk}}$ contained 2 distinct couplings,
$ y_{ai}$ and $y$.
In particular,  if we set  $y_{ai} = 0$ in the Lagrangian then the
symmetry of the Lagrangian was enhanced,  the light neutrino
masses would have been identically zero and the potential for the
$\xi_i$ fields would have also been identically zero.
If $ y_{ai} \neq 0 $ then the  $SU_R(2)$ symmetry is explicitly
broken,  the light neutrino fields pick up their masses as outlined
earlier and also the $\xi_i$'s develop a non-trivial potential.
Furthermore,  since setting  $y_{ai} = 0$ enhances the symmetry
of the Lagrangian,  there is a symmetry which protects the small
parameters in this model.
In the model we are studying the coupling of the light neutrino
fields to the  $\xi$ fields is identically zero if  $y_{ai} = 0$
and arises at the second order in perturbation theory in the
see-saw diagonalization if  $ y_{ai} \neq 0 $.
It is the coupling of the neutrino fields to the  $\xi$ fields
that is responsible for a non-zero effective potential for
the  $\xi$ fields, hence there is a prefactor to the entire effective
potential (including the vacuum energy part) which is proportional
to the appropriate power of $ y_{ai} $.
The observable quantity that $ y_{ai} $ corresponds to is
$ m_{\nu}$ as given in section 3.1 .Since the effective potential has
four mass dimensions, the dimensional prefactor multiplying
the potential ends up being  $ m_{\nu}^4$.
Thus it is natural that the contribution to the vacuum energy density
due to the fields appearing in the see-saw model of neutrino masses
presented here ends up being  $ \sim  m_{\nu}^4$.
One, of course, has to worry about the contribution of other
heavier fields to the vacuum energy density in the cosmological
context. The underlying picture being used is that discussed by
Wilczek\cite{Wilczek}, Reuter and Wetterich\cite{rwet} and
Freese, Adams, Frieman and Mottola\cite{FAFM}.
They argue that the cosmological constant will decay during
the evolution of the universe as the vacuum energy of heavier
fields dissipates due to their coupling to lighter fields.
One would then expect the vacuum energy density at late times
to be dominated by the contribution of the lightest most
weakly coupled fields such as those appearing in the see-saw
model of neutrino masses discussed in this paper. An in-depth
analysis of the details of this mechanism will be the subject of
a later work.



The time evolution of the system can be summarized as follows.  The
fields evolved
to the minimum of their potential on a time scale which is
short compared to the typical
Hubble time scale in the problem.  The evolution of
the scale factor follows the
normal matter dominated behaviour for a while until
the vacuum energy starts playing an
 important role.  After this time the
vacuum energy starts driving the time evolution of
the scale factor. Thus it is the value of the
vacuum energy density that determines the
asymptotic time evolution of the system.
For our model with the choice of
parameters stated and rationalised above we have the final vacuum energy
density $\simeq  m_{\nu}^4$.

The observationally important plot is the
plot of the Hubble parameter as function of
time.  This is displayed in figure 1.  As you can
see the Hubble parameter assumes a
constant value after the vacuum energy starts
playing the dominant role in the evolution
 of the scale factor.

In fact,  in retrospect one can understand the
time evolution of the coupled differential
 equations simply by noting the order of
magnitude of the quantities involved in the
evolution equations.

Let us introduce the following dimensionless physical quantities,
$ \tau = H~t $ ,
$\eta_{i} = \frac{\xi_i}{f} $
and
${\cal{V}} = \frac{V}{m_\nu^4} $
We'll use the following
physical quantities to make the dimensionless quantities of order 1,
$H = 75~h~ km~sec^{-1}~Mpc^{-1}$,  $f = 10^{12}~f_{12} ~ GeV$ and
$m_{\nu} = 2. 5~m_d \times 10^{-3} eV$.

Here are some of the quantities of interest and their magnitudes.  First,  the
expressions
for the pressure density and energy density of the fields is given below:

\be
p_{\eta} = m_{\nu}^4 \left\{ 6. 5 \times 10^{-14} \frac{h^2 f_{12}^2}{m_d^4}
\left[ \left( \frac{d \eta_1}{d \tau} \right)^2 + \left( \frac{d \eta_3}{d
\tau} \right)^2 \right] - {\cal{V}}(\eta_1, \eta_3) \right\}
\ee

\be
\rho_{\eta} = m_{\nu}^4 \left\{ 6. 5 \times 10^{-14} \frac{h^2 f_{12}^2}{m_d^4}
\left[ \left( \frac{d \eta_1}{d \tau} \right)^2 + \left( \frac{d \eta_3}{d
\tau} \right)^2 \right] + {\cal{V}}(\eta_1, \eta_3) \right\}
\ee

Since we have scaled quantities so that
$m_d~, ~h, ~ \eta_i, ~ \tau~ and ~ {\cal{V}}$ are all of order 1 it follows
that,

\be
\rho_{\xi} \simeq V(\xi_1, \xi_3)
\ee
\be
p_{\xi} \simeq -  V(\xi_1, \xi_3)
\ee
Thus in fact,  for all practical purposes,  we have
\be
\rho_{\xi} \simeq - p_{\xi}
\ee
which is the equation of state for vacuum energy and hence demonstrates that
this solution is very close in spirit to the
cosmological constant solution for the
age vs Hubble constant problem.  What we have achieved
is to provide a physical basis
for the correct order of magnitude for this effective cosmological constant.
This can be seen clearly by making the evolution equation for the
scale factor dimensionless too.
\begin{eqnarray*}
\lefteqn{\frac{1}{R(\tau)}\frac{d R(\tau)}{d \tau} + \frac{k}{R^2(\tau)  H^2}
= } \\
& & \Omega_m(\tau_0) \left(\frac{R(\tau_0)}{R(\tau)} \right)^3
+\Omega_r(\tau_0) \left(\frac{R(\tau_0)}{R(\tau)} \right)^4 \\
& &+\left\{\frac{8 \pi}{3} 4. 0 \times 10^{-8} \frac{h^2 f_{12}^2}{m_d^4}
\left[ \left( \frac{d \eta_1}{d \tau} \right)^2 + \left( \frac{d \eta_3}{d
\tau} \right)^2 \right] + 0. 85 \frac{m_d^4}{h^2} {\cal{V}}(\eta_1, \eta_3)
\right\}
\end{eqnarray*}

Note that it is the fact that the mass of the neutrinos
is the correct order of magnitude which allowed the age of the universe
to become compatible with the
 Hubble constant observed today.

The picture presented in this article is of a
vacuum energy that changes as a function
of time due to the coupling of the fields
responsible for the vacuum energy to the other
 fields.  One may worry therefore that the vacuum
energy may disappear because of the
 coupling of the $\xi$ fields to other fields.
However because $\xi$ can only decay
into the lighter neutrinos it is coupled to,
the time scale on which this vacuum energy
 will dacay is much larger than the present age of the universe.
This can be seen
by calculating the decay width of the $\xi$ , $\Gamma_{\xi}$.

This decay width of the $\xi$ particles arises because of the coupling of the
$\xi$
to the lighter fermions with coupling y is given by\cite{rockymike}
\be
\Gamma_{\xi} = \frac{y^2 m_{\xi}}{8 \pi}
\ee
where,
\be
m_{\xi}^2 \simeq \frac{\partial ^2 V}{\partial \xi ^2}
\ee
Therefore,
\be
m_{\xi}^2 \simeq \frac{m_{\nu}^4}{f^2}
\ee
Thus the timescale on which the energy in the
$\xi$ fields is converted into the energy
of $\nu$'s is given by
\be
\Gamma_{\xi}^{-1} \simeq \frac{8 \pi}{y^2 m_{\nu}} \frac{f}{m_{\nu}}
\ee
which is much greater than the present age of the universe.
Thus this vacuum energy is clearly not a short lived thing.
As the expression above displays this is a consequence both of the
fact that we have light particles involved and that they are extremely weakly
coupled to other particles.

In conclusion, the MSW solution to the solar neutrino problem seems to imply a
muon
neutrino mass of a few meV. This in turn would lead to a phase transition in
the PNGB fields associated with massive neutrinos with a critical temperature
of several meV.
This would then give us a vacuum energy density $ \sim (10^{-3}~eV)^4 $,
which would help resolve the conflict between the independent
determinations of the Hubble constant and the age of the universe.
This phase transition also
happens at the correct epoch in the
evolution of the universe to provide a possible explanation of the peak in
quasar space density at redshifts of $2$ to $3$\cite{quasars}.

\newpage

\centerline{\bf ACKNOWLEDGEMENTS}

It's always a  pleasure for me
to thank Richard Holman for many valuable discussions, critical comments and
helpful suggestions.
I would like to thank the organisors of the SLAC
Summer Institute on Particle Physics, Astrophysics and Cosmology
for an extremely stimulating program. I would also like to thank
Michael Turner,  Joel Primack,  Mark Srednicki and  Lincoln Wolfenstein
for stimulating
discussions on this and related topics.
This work was supported in part by the DOE contract DE--FG02--91ER--40682.

\newpage

\vspace{36pt}

\centerline{\bf FIGURE CAPTIONS}

\frenchspacing

{\bf Figure 1 :} Time evolution of the Hubble Parameter in the LTPT model :
$\frac{\dot R(t)}{R(t)}$ Vs.  Time.

\vspace{36pt}

\end{document}